\documentclass{article}
\usepackage{graphicx}
\usepackage{amsmath}
\usepackage{amsfonts}
\usepackage{amssymb}

\begin{document}

\title{Experimental requirements for quantum communication complexity protocols}
\author{E. F. Galv\~{a}o\\Centre for Quantum Computation, Clarendon Laboratory,\\Univ. of Oxford, Oxford OX1 3PU U.K.}
\maketitle
\begin{abstract}
I present a simple two-party quantum communication complexity protocol with
higher success rate than the best possible classical protocol for the same
task. The quantum protocol is shown to be equivalent to a quantum non-locality
test, except that it is not necessary to close the locality loophole. I derive
bounds for the detector efficiency and background count rates necessary for an
experimental implementation and show that they are close to what can be
currently achieved using ion trap technology. I also analyze the requirements
for a three-party protocol and show that they are less demanding than those
for the two-party protocol. The results can be interpreted as sufficient
experimental conditions for quantum non-locality tests using two or three
entangled qubits.
\end{abstract}

\section{Introduction\label{sec intro}}

In the last few years there has been a growing interest in quantum information
theory, in particular in uses of entanglement like teleportation \cite{Bennett
BCJPW 93}, quantum cryptographic key distribution \cite{q cryptography} and
dense coding \cite{Bennett W 92}. In addition to extensive theoretical work,
some tasks involving manipulation of entanglement have been successfully
demonstrated in practice, including teleportation \cite{exp teleport} and
quantum key distribution \cite{exp crypto}. In this paper I discuss the
feasibility of experimentally implementing another application of quantum
entanglement: quantum communication complexity protocols.

The scenario of communication complexity (CC) was introduced by Yao \cite{Yao
79}, who investigated the following problem involving two separated parties
(Alice and Bob). Alice receives a $n$-bit string $x$ and Bob another $n$-bit
string $y$, and the goal is for one of them (say Bob) to compute a certain
function $f\left(  x,y\right)  $ with the least amount of communication
between them. Of course they can always succeed by having Alice send her whole
$n$-bit string to Bob, who then computes the function, but the idea here is to
find clever ways of calculating $f$ with less than $n$ bits of communication.
This problem is relevant in many contexts: in electronic circuit design, for
example, one wants to minimize energy use by decreasing the amount of electric
signals required between the different components during a distributed computation.

Various authors have established \cite{Cleve B 97,Buhrman CvD 97,Buhrman vDHT
99,Raz 99,Hardy vD 99} the somewhat surprising result that entanglement can be
used to reduce the amount of communication necessary to calculate many
functions of distributed inputs. This can be done by previously sharing
entangled states between the parties, and then allowing them to do
measurements on these states as part of their CC protocols. Quantum CC tries
to quantify the gain obtained by using entanglement for different classes of
multi-party distributed function calculations.

In this paper I present a simple two-party CC task and discuss some necessary
conditions to demonstrate the quantum protocol in the laboratory. I start by
describing the task itself in section \ref{sec task} and its optimal classical
protocol (section \ref{sec class prot}), followed by a more efficient quantum
protocol (section \ref{sec quant prot}). In section \ref{sec bell} I show that
the efficiency of the quantum protocol relies on its equivalence to
measurements that maximally violate the Clauser-Horne-Shimony-Holt (CHSH)
quantum non-locality inequality. In section \ref{sec experim}\ I discuss some
issues concerning the feasibility of implementing this protocol
experimentally, and argue that ion trap technology is presently very close to
achieving the necessary thresholds. In section \ref{sec three party} I analyse
a three-party CC task discussed in \cite{Buhrman CvD 97} and show that in
principle it can be demonstrated experimentally with lower detector
efficiencies and higher background count rates than the two-party quantum CC
protocol previously discussed. Finally in section \ref{sec conclu} I round up
with some concluding remarks.

\section{A two-party communication complexity task\label{sec task}}

In \cite{Hardy vD 99} the authors presented a two-party CC task on which I
will now elaborate, introducing a slight change in order to obtain a larger
gap in efficiency between the quantum and the classical protocols. This larger
difference will be important when we discuss the experimental implementation
in section \ref{sec experim}, as it allows for lower detector efficiencies.

The modified task can be simply stated as follows. The two parties Alice and
Bob are each given a number between $0$ and $2N-1$, where $N$ is an even
integer and $N\geq4$. We can think of Alice's number $x$ and Bob's number $y$
as being integers modulo $2N$ lying uniformly distributed on a circle. The
numbers are chosen randomly with equal probabilities, but obeying certain
constraints: either%

\begin{equation}
\text{a) }(x-y)\operatorname{mod}2N\in\{1,2N-1\}, \label{constraint a}%
\end{equation}
in which case $x$ and $y$ are said to be `neighbours', or%

\begin{equation}
\text{b) }(x-y)\operatorname{mod}2N\in\{N-1,N+1\}, \label{constraint b}%
\end{equation}
in which case $x$ and $y$ are `anti-neighbours'.

The reason for the terms `neighbours' and `anti-neighbours' is obvious when
one looks at the two possibilities in the integers modulo $2N$ circle: in the
`neighbours' case $x$ and $y$ are adjacent points; in the `anti-neighbours'
case $y$ is adjacent to the farthest point from $x$, which is
$(x+N)\operatorname{mod}2N$. After being assigned their numbers, Alice is then
allowed to communicate a single bit of information to Bob, who then has to
decide whether $x$ and $y$ are neighbours or anti-neighbours, achieving as low
an error rate as possible over many runs of the task.

This protocol represents only a slight departure from that presented in
\cite{Hardy vD 99}. The difference here is that $x$ and $y$ are not allowed to
coincide or to differ by $N(\operatorname{mod}2N)$, as was the case in
\cite{Hardy vD 99}. As we will see in section \ref{sec quant prot}, this
eliminates two possibilities for which the quantum protocol does not offer
advantage over the best classical one, making the quantum protocol more
efficient in relation to the optimal classical counterpart.

\subsection{The best possible classical protocol\label{sec class prot}}

In this section I describe the best possible deterministic classical protocol,
and argue that it is optimal. After receiving her number $x$, Alice decides
about which bit value to send to Bob by looking up the value of some function
$g\left(  x\right)  =\pm1$, as $x$ is only data she has access to. The
function $g\left(  x\right)  $ is also known to Bob, who upon receiving the
bit from Alice learns that $x$ lies in one of two disjunct sets, and acts in
an optimal way to try to decide whether $x$ and $y$ are neighbours or antineighbours.

We may think of the function $g(x)$ as a colouring of the $2N$ points in the
circle in either black (corresponding to $g(x)=+1$) or white ($g(x)=-1$). What
Alice does is to send $x$'s colour to Bob for him to decide which case they
have. Suppose, for example, that Alice sends the bit $+1$ to Bob,
corresponding to a black $x$. If Bob's $y$ has two black neighbours and two
white anti-neighbours, then Bob can be absolutely sure that $x$ and $y$ are
neighbours. Of course the situation will not always be so clear-cut, forcing
Bob to make informed or random guesses sometimes. One way to find the optimal
deterministic classical protocol is to consider all possible $g(x)$
(=colourings) and evaluate the probability of success for Bob, using the fact
that the distributions for $x$ and $y$ are uniform. It is a reasonable
conjecture that one optimal colouring consists of one half of the circle
coloured white and the other half black, which yields a success probability of%

\begin{equation}
p_{c}^{N}=\frac{N-1}{N}. \label{pc N}%
\end{equation}
For $N=4$, I analysed the 16 non-equivalent possible colourings and indeed
found that the optimal deterministic protocol is of the form above, having a
success probability of%
\begin{equation}
p_{c}^{N=4}=\frac{3}{4}. \label{pc 4}%
\end{equation}
Even though we considered only deterministic protocols, in \cite{Hardy vD 99}
the authors showed that probabilistic protocols need not be considered for
this kind of problem. This is true because any probabilistic protocol can be
reduced to a deterministic one by having Alice and Bob share random numbers
before they are assigned $x$ and $y$. As a result, for $N=4$ the best possible
classical protocol is the one described above, achieving a success probability
of $p_{c}=3/4$.

\subsection{The quantum protocol\label{sec quant prot}}

Here I describe a quantum protocol which accomplishes the task with a larger
success probability than the optimal classical protocol described above. The
motivation for this protocol comes from the visualization of $x$ and $y$ on
the $\operatorname{mod}2N$ circle, where they lie separated by an angle which
is either $\pi/N$ or $(N-1)\pi/N$. It is a well known fact that local rotation
followed by measurements on maximally entangled states yield strong
correlations which are not allowed classically \cite{Bell 65,Clauser HSH
69,Hardy 91}, and which will be used in the protocol.

Before they are given their numbers, Alice and Bob meet and share a singlet state%

\begin{equation}
\left|  \psi\right\rangle =\frac{1}{\sqrt{2}}\left(  \left|  +\right\rangle
_{A}\left|  -\right\rangle _{B}-\left|  -\right\rangle _{A}\left|
+\right\rangle _{B}\right)  . \label{singlet}%
\end{equation}
For simplicity of description, we will assume that $\left|  \psi\right\rangle
$ was shared in the form of an entangled pair of spin-$1/2$ particles. Upon
receiving her number, Alice measures the spin on her particle along the axis
defined by the angle $\theta=\pi x/N$ to the $z$ axis on the $xy$ plane, and
sends the result ($\pm1$) to Bob. He measures the spin on his particle along
the axis defined by the angle $\phi=\pi y/N$ to the $z$ axis on the $xy$
plane, also obtaining a result $\pm1$. It does not matter whether he does this
before or after receiving Alice's message. Bob's decision is made in a simple
way: if the two measurement results are the same he guesses that $x$ and $y$
are neighbours, otherwise that they are anti-neighbours.

Let us now evaluate the probability of error if Bob and Alice adopt this
protocol. For the singlet state one can easily show that%
\begin{align}
p(\text{same}  &  \mid\theta,\phi)=\frac{1}{2}(1-\cos(\theta-\phi
))\label{psame}\\
p(\text{opposite}  &  \mid\theta,\phi)=\frac{1}{2}(1+\cos(\theta-\phi)),
\label{popposite}%
\end{align}
where $p($same$\mid\theta,\phi)$ is the probability that their outcomes are
the same and $p($opposite$\mid\theta,\phi)$ is the probability that their
outcomes are opposite. In the `neighbours' case we have $\left|  \theta
-\phi\right|  =\pi/N$ and hence%

\[
p_{\text{success}}(\text{neighbours})=p(\text{opposite}\mid\theta,\phi
)=\frac{1}{2}(1+\cos(\pi/N)).
\]
In the `anti-neighbours' case we have either $\left|  \theta-\phi\right|
=\pi+\pi/N$ or $\left|  \theta-\phi\right|  =\pi-\pi/N$, from which we find%

\[
p_{\text{success}}(\text{anti-neighbours})=p(\text{opposite}\mid\theta
,\phi)=\frac{1}{2}(1+\cos(\pi/N)).
\]

In any case, we see that this protocol is successful with a probability%

\begin{equation}
p_{q}=\frac{1}{2}(1+\cos(\pi/N)),
\end{equation}
which is always larger than the conjectured classical success probability
given by eq. (\ref{pc N}), and definitely larger than the best classical
protocol for $N=4$, whose success probability is limited to $p_{c}=3/4$
[compare with $p_{q}=1/2(1+\cos(\pi/4))\simeq0.8536$].

Unlike the classical protocol, the quantum version has a probability of
success which is independent of the probability distribution for $x$ and $y$.
This, however, should not be considered a substantial advantage for the
quantum protocol, as a randomized classical protocol can be used to obtain
$p_{c}=(N-1)/N$ even for non-uniform distributions of $x$ and $y$. This can be
done by having Alice and Bob choose different optimal colourings for each run,
according to previously shared random numbers.

In the protocol described above the singlet state (\ref{singlet}) is measured
along directions that differ by either $\pi/N$ or $(N-1)\pi/N$. This is in
contrast with the quantum protocol for the task presented in \cite{Hardy vD
99}, whose measurements include those but also measuments along the same axis.
Since the quantum and classical correlations are equal (and maximal) for
measurements along the same axis, we obtain a larger gap between the classical
and quantum efficiencies by eliminating this possibility, which we did by
redefining the task suitably.

\subsection{Quantum protocol as a Bell inequality test\label{sec bell}}

The quantum protocol described above is clearly inspired by the CHSH
inequality presented in \cite{Clauser HSH 69}, which establishes bounds for
classical correlations between simultaneous measurements on maximally
entangled states. In fact, we will see that there is a simple interpretation
of the quantum protocol in terms of Bell-type quantum non-locality tests.

In \cite{Hardy 91} an inequality was derived for classical correlations
between measurements with two possible outcomes (denoted by $+$ or $-$),
arising from experiments with $N/2$ different setups at Alice's laboratory and
$N/2$ at Bob's. We write $p^{e}(a_{i},b_{j})$ to denote the probability of
obtaining two equal outcomes ($++$ or $--$) with setups $a_{i}$ at Alice and
$b_{j}$ at Bob, and similarly $p^{d}(a_{i},b_{j})$ for different outcomes
($+-$ or $-+$). For $N=4$ we have the following inequality \cite{Hardy 91}:%

\begin{equation}
p^{d}(a_{1},b_{1})+p^{d}(a_{2},b_{1})+p^{d}(a_{2},b_{2})+p^{e}(a_{1}%
,b_{2})\leq3, \label{hardy ineq.}%
\end{equation}
which must be obeyed by \textit{any} local theory. This inequality is
saturated by a simple `classical spin' local model described in \cite{Bell
65,Hardy 91,Peres 93}. It is easy to see that quantum mechanics violates
inequality (\ref{hardy ineq.}) for measurements on the singlet state
(\ref{singlet}). Choosing the angles of measurement to be $a_{1}=0,a_{2}%
=\pi/2,b_{1}=\pi/4,b_{2}=3\pi/4$ we obtain maximal violation of inequality
(\ref{hardy ineq.}):%

\begin{align}
p_{q}^{d}(a_{1},b_{1})  &  =p_{q}^{d}(a_{2},b_{1})=p_{q}^{d}(a_{2}%
,b_{2})=p_{q}^{e}(a_{1},b_{2})=\frac{1}{2}\left(  1+\frac{\sqrt{2}}{2}\right)
\label{prob ineq}\\
&  \Rightarrow p_{q}^{d}(a_{1},b_{1})+p_{q}^{d}(a_{2},b_{1})+p_{q}^{d}%
(a_{2},b_{2})+p_{q}^{e}(a_{1},b_{2})=2+\sqrt{2}>3. \label{ineq viol}%
\end{align}
We see that the quantum mechanics predictions violate the inequality, and
therefore are impossible to be accounted for by any local theory. The same
setup also maximally violates the CHSH inequality, as discussed in
\cite{Clauser HSH 69}; in fact Hardy has shown \cite{Hardy 91} that the CHSH
inequality can be obtained from (\ref{hardy ineq.}).

It is interesting to note that there is an exact correspondence between the
measurements that maximally violate inequality (\ref{hardy ineq.}) and the
quantum protocol for $N=4$ described above. In both cases the two parties do
measurements on a singlet state along co-planar directions separated by two
possible angles: $\pi/4$ or $3\pi/4$ radians. In the quantum protocol we want
to maximize%

\[
p_{q}=p_{success}(\text{neighbours})+p_{success}(\text{anti-neighbours}).
\]
The inequality test described above corresponds to a similar situation.
Quantum mechanics predicts $p_{q}^{d}(a_{1},b_{1})=p_{q}^{d}(a_{2}%
,b_{1})=p_{q}^{d}(a_{2},b_{2})=$ $p_{success}($neighbours$)$ and $p_{q}%
^{e}(a_{1},b_{2})=p_{success}($anti-neighbours$)$, and we maximize
$3p_{success}($neighbours$)+$ $p_{success}($anti-neighbours$)$ in order to
violate the inequality by the greatest amount$.$ As $p_{success}%
($neighbours$)=p_{success}($anti-neighbours$)$, the maximization of the
violation of inequality (\ref{ineq viol}) also maximizes the probability of
success $p_{q}$. Curiously, the `classical spin' model used to saturate the
inequality turns out to be equivalent to the best classical protocol for the
task, that is, it can be used as an alternative to the optimal classical
protocol described in section \ref{sec class prot}.

The key difference between the measurements used in the quantum protocol and
the CHSH inequality test is the number of possible setups at Alice and Bob.
For the CHSH inequality violation we consider two setups at Alice and two at
Bob, whereas in the quantum protocol Alice and Bob need to use one of four
possible setups. The `promise' that $x$ and $y$ are either neighbours or
antineighbours guarantees that for each run the measurements at Alice and Bob
will be done along angles that differ either by $\pi/4$ or $3\pi/4$ radians,
like the CHSH test. But unlike it, the quantum protocol relies on Alice using
one of four possible setups; she cannot communicate this to Bob with a single
bit. Instead, she communicates the result of the measurement to Bob, who can
then use the stronger-than-classical correlations between the measurements to
obtain success with a high probability.

\subsection{Experimental implementation\label{sec experim}}

The quantum CC protocol of section \ref{sec quant prot} with $N=4$ is arguably
the simplest to implement in the laboratory. There is a clear-cut criterion
for experimental success: if after a large number of runs the success rate is
found to be larger than $3/4$ then entanglement-enhanced CC will have been
demonstrated. In \cite{Buhrman CvD 97} a similar CC task was presented which
has the same classical and quantum error rates as the one here, and which
therefore needs an equivalent detector efficiency to be implemented (as we
will see below). The advantages of the present task is that it is simpler to
describe and easier to interpret in terms of Bell-type inequalities, as
discussed in the previous section.

It is important to point out that in order to experimentally implement this
quantum CC protocol it is \textit{not} necessary to have space-like separated
measurements at Alice and Bob. Unlike Bell inequality tests, here the two
parties are \textit{required} to communicate before finishing the task. All
that is needed for a successful experiment is to implement the quantum
protocol in such a way as to obtain a success rate which is higher than that
of the highest classical protocol. Once this is done, one might correctly
argue that, from a fundamental point of view, local hidden-variable theories
cannot be ruled out as the responsible for the better performance of the
quantum protocol. From a practical point of view, however, after a large
enough number of runs this would be an undeniable demonstration of
entanglement-assisted communication. The question of whether local hidden
variable theory or quantum mechanics is the correct theory to explain the
enhancement becomes immaterial if we are only concerned about establishing an
experimental evidence for the effect.

Given the equivalence of the protocol with a CHSH inequality test, the
experimental problems to be taken into account are those of a CHSH test,
except that we do not need to close the locality loophole. With this in mind,
we can calculate necessary bounds for the detector efficiency and background levels.

For the quantum protocol to beat the best classical one for $N=4$ we need a
success rate higher than $p_{c}=3/4$. First I will analyse the case of no
background detections, and detectors that either obtain the correct result or
fail to register the event \cite{vDam 99}. With a finite detector efficiency
$\eta<1$, Alice and Bob must agree on a procedure for the case when their
detectors fail. They are not allowed to communicate the failure, as this would
consist of further bits of communication between the parties, and the
communication allowed is restricted to one bit. The most effective way is for
Alice to proceed with the best classical protocol in case her detector fails,
and for Bob to do the same when his detector fails. Whenever both detectors
fail, Alice and Bob will obtain the best classical sucess rate. In the case of
a single detector failure, Bob's guess will still be correct with $p=1/2$.
Assuming independent errors at the two detectors, we can calculate the minimum
detector efficiency needed:%

\begin{align*}
&  \eta^{2}p_{q}+(1-\eta)^{2}p_{c}+\eta(1-\eta)\frac{1}{2}+(1-\eta)\eta
\frac{1}{2}>p_{c}\\
&  \Rightarrow\eta>\frac{2p_{c}-1}{p_{q}+p_{c}-1}.
\end{align*}
For the case $N=4$ that we analysed, we have $p_{q}=1/2+\sqrt{2}/4$ and
$p_{c}=3/4$, which results in a minimum necessary detector efficiency%

\[
\eta_{\min}=2(\sqrt{2}-1)\simeq0.828,
\]
which is also the minimum detector efficiency necessary for a loophole-free
CHSH inequality test \cite{Garg M 87}. A similar analysis can be done for
general $N$, and it turns out that the minimum detector efficiency required
increases with increasing $N$, making the $N=4$ case the most interesting for
an experimental test.

We can include background detections in the analysis by introducing a factor
$\mu$, defined such that $(1-\mu)$ is the fraction of detections that are due
to background photons. We assume completely random background detections,
yielding each measurement outcome with probability $p=1/2$. Note that
imperfections in the state preparation procedure can also be incorporated in
the model through the parameter $\mu$. Each run of the experiment can then be
classified into one of three categories:

1) Alice and Bob measure their parts of the EPR pair accurately with
probability $\eta^{2}\mu^{2}$, in which case they apply the quantum protocol
and succeed with probability $p_{q}$.

2) Both their detectors fail to detect anything, which will happen in a
fraction $(1-\eta)^{2}$ of the runs. In this situation they will both use the
classical protocol with a probability of success $p_{c}$.

3) In all the other situations Bob will use either the quantum or the
classical protocol, depending on whether he detects a photon or not. However,
due to either background or lack of detection at Alice's side his guesses will
be random, yielding a success rate of only $1/2$.

Taking all this into account, the condition for a higher-than-classical
success rate is:%

\begin{equation}
\eta^{2}\mu^{2}p_{q}+(1-\eta)^{2}p_{c}+(1-\eta^{2}\mu^{2}-(1-\eta)^{2}%
)\frac{1}{2}>p_{c}. \label{mu and eta}%
\end{equation}
With $p_{c}$ and $p_{q}$ from our task with $N=4$, this is equivalent to%

\begin{equation}
\mu>\frac{1}{2\eta}\sqrt{2\sqrt{2}\eta(2-\eta)}. \label{bound visibility}%
\end{equation}
In particular, for perfect detectors ($\eta=1$), we need $\mu>2^{-1/4}%
\simeq0.841$. In Figure 1 we represent the region in the $\eta-\mu$ parameter
space that guarantees a higher-than-classical success rate.

The favourable $\eta-\mu$ region represents exactly the same conditions
necessary for a violation of the CHSH inequality (as can be gathered from
\cite{Larsson99}, for example). This is not a surprise, given the equivalence
between the quantum protocol and a CHSH test pointed out in section \ref{sec
bell}. More generally, whenever we have a quantum CC protocol outperforming
the optimal classical protocol for a given CC task, the measurements involved
in the quantum protocol can be re-interpreted in terms of a non-locality
proof. This is the case at least for tasks allowing only one round of
communication between the parties, as the one presented above and the one we
will analyse in the next section. For this class of CC tasks, it is always
possible to carry out the protocol in such a way as to enforce that all
measurements on entangled states be performed in space-like separated regions;
if despite this, the CC task is performed with higher than classical
efficiency, it must be because of stronger-than-classical correlations among
the results of the measurements of the quantum protocol.

Maximally entangled pairs of photons can be generated with high fidelity and
measured with precision, which guarantees the equivalent of a very high $\mu$.
However, detector efficiencies $\eta$ are still short of those necessary. Ion
traps techniques, on the other hand, can reach high $\eta$ and $\mu$
\cite{Wineland}. This makes ion traps a natural choice for the experimental
demonstration of the simple quantum communication complexity task presented here.

\section{A three-party task\label{sec three party}}

In this section I will discuss the three-party communication complexity (CC)
task described in \cite{Buhrman CvD 97}, finding the necessary experimental
requirements for the demonstration of a quantum protocol for it. The
enhancement in performance with relation to the best classical protocol arises
from measurements on a tripartite Greeberger-Horne-Zeilinger (GHZ) state
\cite{Greenberger HZ 89,Mermin 90,Greenberger HSZ 90}%

\[
\left|  GHZ\right\rangle =\frac{1}{\sqrt{3}}(\left|  0_{A}0_{B}0_{C}%
\right\rangle +\left|  1_{A}1_{B}1_{C}\right\rangle ).
\]
The state preparation and measurement tend to be much harder to perform
experimentally than for the two-party protocol described in section \ref{sec
quant prot}. In principle, however, we will see that the three-party quantum
protocol requires lower detector efficiencies and tolerates higher background
count rates than the two-party protocol discussed above.

The CC task is defined as follows \cite{Buhrman CvD 97}. Alice, Bob and Claire
receive respectively the numbers $x$, $y$ and $z$ $\in U=\{0,1,2,3\}$, and
these are guaranteed to satisfy%

\begin{equation}
(x+y+z)\operatorname{mod}2=0. \label{cond gmn}%
\end{equation}

The goal is for Alice to learn the value of the function%

\[
f(x,y,z)=\frac{1}{2}[(x+y+z)\operatorname{mod}4]
\]
(which can be either $0$ or $1$), after receiving a single bit of
communication from each Bob and Claire. We assume an uniform probability
distribution for $x$, $y$ and $z$, subject to satisfying eq. (\ref{cond gmn}).

There is a quantum protocol which succeeds with probability $p_{q}=1$ (see
\cite{Buhrman CvD 97}). It is inspired by the GHZ non-locality proof of
\cite{Greenberger HZ 89,Mermin 90,Greenberger HSZ 90}.

The best classical protocol, however, succeeds only with a lower probability
$p_{c}$. In order to calculate it, we first note that for each possible $x$
all four possibilities for $y$ and $z$ are allowed by condition (\ref{cond
gmn}), provided they are properly paired. This means that by knowing $x$,
Alice cannot infer any information about the values of Bob's $y$ and Claire's
$z$, as they are all equally likely candidates. Therefore the best possible
procedure for Bob will be to agree beforehand on the encoding scheme for his
message, so that his bit of communication informs Alice that $y$ lies in one
of two disjunct sets, each containing two possible $y$ values. The same holds
for Claire and her one-bit message about $z$.

The task involves the assumption that all $x,y$ and $z$ values satisfying
(\ref{cond gmn}) will be chosen with equal probabilities. As all $16$ possible
combinations of $y$ and $z$ appear the same number of times in the full list
of possibilities, this means that the choice of two-element disjunct sets for
the encoding is not important, all yielding the same probability of success
$p_{c}$. We can calculate $p_{c}$ by choosing any such encoding for Bob and
Claire, and averaging over the probability of success for all possible $x,y$
and $z$. This results in the highest classical success probability%

\[
p_{c}=3/4.
\]

If we are to implement the quantum protocol in the laboratory, then we have to
account for imperfect detectors and state preparation. The analysis is similar
to that presented in section \ref{sec experim}, with the difference that here
we have three detectors instead of two. We define the background rate
$(1-\mu)$ as the fraction of events detected at each detector which are due to
noise; we assume the worst-case scenario of completely random detection
outcomes from these events. The single detector efficiency $\eta$ is the
fraction of all events that result in a detector firing. Whenever we detect a
signal from the three detectors (which will happen with probability $\eta
^{3}\mu^{3}$), the quantum protocol works with probability $p_{q}=1$. If
Alice's detectors does not fire, she will rely on the possibility of Bob's and
Claire's detectors not firing either, and on them using the optimal classical
protocol; this will happen with probability $(1-\eta)^{3}$ and result in a
probability of success $p_{c}=3/4$. In all other cases, Alice will only be
able to make random guesses, succeeding with probability $1/2$. Assuming
independent errors at each detector, for a higher-than-classical probability
of success we need%

\begin{equation}
\eta^{3}\mu^{3}p_{q}+(1-\eta)^{3}p_{c}+[1-\eta^{3}\mu^{3}-(1-\eta)^{3}%
]\frac{1}{2}>p_{c}. \label{ineq tri}%
\end{equation}
Substituting $p_{c}=3/4$ and $p_{q}=1$ into the inequality above, we obtain%

\begin{equation}
\mu>\frac{1}{2\eta}(4\eta^{3}-12\eta^{2}+12\eta)^{1/3}. \label{mueta tri}%
\end{equation}
This inequality gives us the region in the $\mu-\eta$ parameter space that
allows for a successful experimental test of the quantum protocol. For perfect
detectors ($\eta=1$) we need%

\[
\mu>2^{-1/3}\simeq0.794,
\]
whereas for zero background and perfect state preparation and measurement
($\mu=1$), it is sufficient to have%

\[
\eta>\frac{1}{2}(\sqrt{21}-3)\simeq0.791.
\]
Note that these are less demanding requirements than those necessary for the
bipartite quantum CC protocol [see eqs. (\ref{mu and eta}) and (\ref{bound
visibility})]. This is related to the fact that quantum non-locality tests for
multiparticle states can be done with lower detector efficiencies than for the
two-particle case \cite{Larsson 98a,Larsson 98b,Larsson S 00}. Given the
possibility of interpreting the higher-than-classical performance of the
quantum protocol in terms of a quantum non-locality test (as was pointed out
at the end of section \ref{sec experim}), we can view requirement (\ref{mueta
tri}) as a sufficient condition for testing quantum non-locality for three
maximally entangled particles. Besides limited detector efficiency, here we
consider also limited visibility (due to background counts), a situation which
has not been considered before. However, the bound (\ref{mueta tri}) is not
strict; the simplest way to see this is by noting that with $\mu=1$, a GHZ
test can be performed whenever $\eta>.75$ if we assume independent errors
\cite{Larsson 98b}.

\section{Conclusion\label{sec conclu}}

I have presented a simple two-party communication complexity task which can be
implemented with a higher-than-classical success rate using quantum
entanglement. The efficiency of the quantum protocol was shown to be linked to
its equivalence to measurements that maximally violate the CHSH inequality. In
order to obtain a higher success rate than classically possible, it is
sufficient to have detector efficiencies and background count rates compatible
with those necessary for a CHSH test (see Fig. 1).

I have also analysed the experimental requirements for a three-party quantum
communication complexity protocol presented in \cite{Buhrman CvD 97} and have
shown that the requirements are less demanding than for the two-party protocol
discussed. At least for the two-party protocol, the experimental requirements
are within reach of currently available ion trap technology, indicating the
feasibility of demonstrating quantum communication complexity protocols
experimentally in the near future.

Some relations were pointed out between quantum non-locality proofs and
quantum communication complexity protocols. In particular, I have shown that
sufficient conditions for violation of quantum non-locality inequalities can
be derived from the analysis of quantum communication complexity protocols of
two or more parties.

\section{Acknowledgments}

I would like to thank Jan-\AA ke Larsson for pointing out a mistake in an
earlier version of this paper and Lucien Hardy for valuable discussions. I
acknowledge support from the U.K. Overseas Research Studentships scheme and
from the Brazilian agency Coordena\c{c}\~{a}o de Aperfei\c{c}oamento de
Pessoal de Nivel Superior (CAPES).

%

\begin{figure}
[ptb]
\begin{center}
\includegraphics[
height=4.062in,
width=4.7504in
]%
{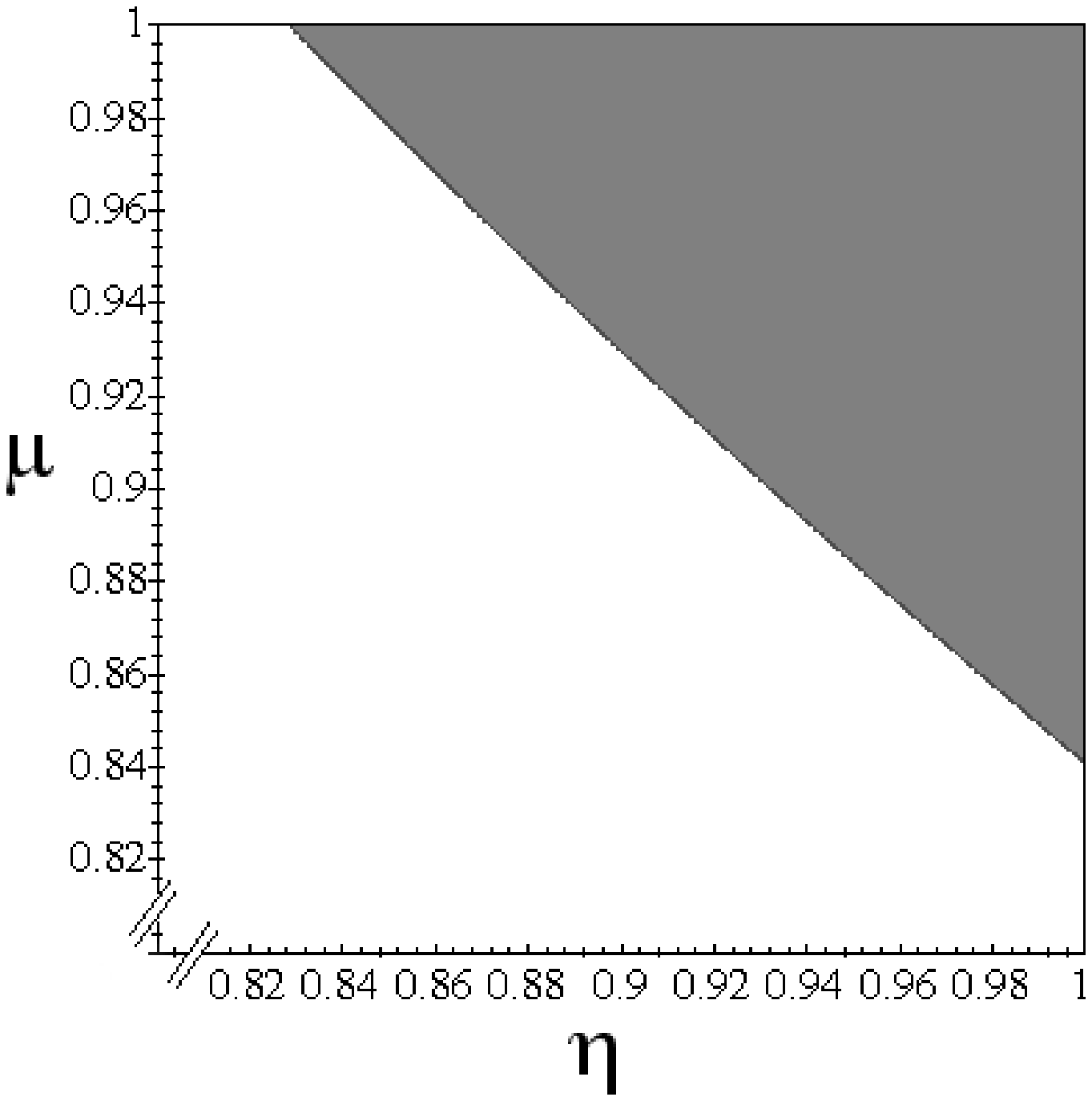}%
\caption{The shaded area indicates the region where the background level
$(1-\mu)$ and detector efficiency $\eta$ allow for a two-party quantum
communication complexity protocol which is more efficient than any classical
one for the same task. The area corresponds to that given by inequality
(\ref{bound visibility}).}%
\end{center}
\end{figure}

\begin{thebibliography}{99}
\bibitem{Bennett BCJPW 93}C. H. Bennett \textit{et al.}, Phys. Rev. Lett.
\textbf{70}, 1895 (1993).

\bibitem {q cryptography}C. H. Bennett and G. Brassard ``Quantum Cryptography:
Public Key Distribution and Coin Tossing'', Proceedings of IEEE International
Conference on Computer Systems and Signal Processing, Bangalore, India,
December 1984, pp 175-179; D. Deutsch \textit{et al.}, Phys. Rev. Lett.
\textbf{77}, 2818 (1996); \textit{erratum} \textbf{80}, 2022 (1998); D.
Mayers, LANL e-print quant-ph/9802025; H. -K. Lo, H. F. Chau, LANL e-print quant-ph/9803006.

\bibitem {Bennett W 92}C. H. Bennett, S. J. Wiesner, Phys. Rev. Lett.
\textbf{69}, 2881 (1992).

\bibitem {exp teleport}D. Bouwmeester \textit{et al.}, Nature (London)
\textbf{390}, 575 (1997); D. Boschi \textit{et al.,} Phys.Rev.Lett.
\textbf{80}, 1121 (1998); A. Furusawa \textit{et al.}, Science \textbf{282},
706 (1998).

\bibitem {exp crypto}C. H. Bennett et al.,Lecture Notes in Computer Science
\textbf{473}, 253 (1990); C. H. Bennet et al., J. Cryptology \textbf{5}, 3
(1992); B. C. Jacobs, J. D. Franson, Opt. Lett. \textbf{21}, 1854 (1996).

\bibitem {Yao 79}A. C. Yao, in Proc.11th Ann. ACM Symp. on Theory of Computing
(1979), pp. 209-213.

\bibitem {Cleve B 97}R. Cleve and H. Buhrman, Phys. Rev. A \textbf{56}, 1201 (1997).

\bibitem {Buhrman CvD 97}H. Buhrman, R. Cleve and W. van Dam, Los Alamos
electronic preprint quant-ph/9705033.

\bibitem {Buhrman vDHT 99}H. Buhrman \textit{et al.}, Phys. Rev. A
\textbf{60}, 2737 (1999).

\bibitem {Raz 99}R. Raz, Proc. 31st Ann. ACM Symp. on Theory of Computing
(1999), pp. 358-367 .

\bibitem {Hardy vD 99}L. Hardy and W. van Dam, Phys. Rev. A \textbf{59}, 2635 (1999).

\bibitem {Bell 65}J. S. Bell, Physics \textbf{1}, 195 (1965).

\bibitem {Clauser HSH 69}J. F. Clauser \textit{et al.}, Phys. Rev. Lett.
\textbf{23}, 880 (1969).

\bibitem {Hardy 91}L. Hardy, Phys. Lett. A \textbf{161}, 21 (1991).

\bibitem {Peres 93}A. Peres, \textit{Quantum theory: concepts and methods}
(Kluwer Academic, Dordrecht 1993), chapter 5.

\bibitem {vDam 99}This calculation was also done independently by W. van Dam,
\textit{Non-locality \& Communication Complexity}, PhD Thesis, Department of
Physics, University of Oxford (1999).

\bibitem {Garg M 87}A. Garg, N. D. Mermin, Phys. Rev. D \textbf{35}, 3831 (1987).

\bibitem {Wineland}Q. A. Turchette \textit{et al.}, Phys. Rev. Lett.
\textbf{81}, 3631 (1998); C. Monroe \textit{et al.}, `Scalable entanglement of
trapped ions', to appear in Proc. 17th Int. Conf. on Atomic Physics; D.
Wineland, private communication (2000).

\bibitem {Larsson99}J.-\AA. Larsson, Phys. Lett. A\textbf{ 256}, 245 (1999).

\bibitem {Greenberger HZ 89}D. M. Greenberger, M. A. Horne, and A. Zeilinger,
in \textit{Bell's Theorem, Quantum Theory, and Conceptions of the Universe},
edited by M. Kafatos (Kluwer Academic, Dordrecht, 1989).

\bibitem {Mermin 90}N. D. Mermin, Phys. Today \textbf{43} (6), 9 (1990).

\bibitem {Greenberger HSZ 90}D. M. Greenberber, M. A. Horne, A. Shimony, and
a. Zeilinger, Am. J. Phys. \textbf{58}, 1131 (1990).

\bibitem {Larsson 98a}J.-\AA. Larsson, Phys. Rev. A \textbf{57}, 3304(1998).

\bibitem {Larsson 98b}J.-\AA. Larsson, Phys. Rev. A \textbf{57}, R3145 (1998).

\bibitem {Larsson S 00}J.-\AA. Larsson and J. Semitecolos, Los Alamos
electronic preprint quant-ph/0006022.
\end{thebibliography}
\end{document}